\newif\ifdraft
\draftfalse
\newif\ifcameraready
\camerareadytrue

\documentclass[camera,letterpaper,nomarginnotes,nonarrowgutter]{jpaper}

\usepackage[bookmarks=true,breaklinks=true,colorlinks,linkcolor=black,citecolor=NavyBlue,urlcolor=blue]{hyperref}
\usepackage{algorithmic}
\usepackage{graphicx}
\usepackage{textcomp}
\usepackage{xcolor}
\usepackage[flushleft]{threeparttable}

\usepackage{titlecaps}

\usepackage{balance}
\usepackage{textcomp}
\usepackage{pifont}
\usepackage{booktabs}
\usepackage{glossaries} % For acronyms / abbreviations
\setkeys{glslink}{hyper=false} % to disable hyperlink for glossaries
\usepackage{setspace}
\usepackage{listings}
\usepackage[linesnumbered]{algorithm2e}
\usepackage{siunitx} % To use SI unit system 
\usepackage{subcaption}
\usepackage{tikz}
\usepackage[dvipsnames]{xcolor}
\usepackage{multirow}
\usepackage{enumitem}
\usepackage[utf8]{inputenc}

\usepackage{titlesec}

\usepackage{interval} %sec5, interval
\usepackage{duckuments} %duck figures
\usepackage{algorithmic} %algoritmic

\usepackage[us,12hr]{datetime}
\usepackage[en-GB, useregional=numeric]{datetime2}
\usepackage{fourier-orns}
\usepackage{fancyhdr}
\usepackage{pgfornament}
\usetikzlibrary{calc}
\usepackage{url}
\usepackage{yfonts}
\usepackage[sort,compress]{cite}

\usepackage{mathptmx} % This is Times font

\definecolor{gfored}{rgb}{0.580, 0.050, 0.211}
\definecolor{ao}{rgb}{0.007, 0.520, 0.867}
\definecolor{moegi}{rgb}{0.357, 0.537, 0.188}
\definecolor{jl}{rgb}{1.0, 0.2, 0.8}
\definecolor{brown(web)}{rgb}{0.65, 0.16, 0.16}
\definecolor{bisque}{rgb}{1.0, 0.89, 0.77}
\definecolor{nbs}{rgb}{0.88, 0.07, 0.37}
\definecolor{yt}{rgb}{0.58, 0.44, 0.86}
\definecolor{iy}{rgb}{0.0, 0.36, 0.05}
\definecolor{ouscolor}{rgb}{0.0, 0.2, 0.4}
\newcommand{\dingOne}{\circledtest{1}}
\newcommand{\dingTwo}{\circledtest{2}}
\newcommand{\dingThree}{\circledtest{3}}

\newcommand{\one}{1)}
\newcommand{\two}{2)}
\newcommand{\three}{3)}
\newcommand{\four}{4)}

\newcommand{\xxx}[1]{\param{XXX}} % to highlight hardcoded numbers

\newcommand{\ignore}[1]{}

\ifdraft
    \usepackage[colorinlistoftodos,prependcaption,textsize=tiny]{todonotes}
    \usepackage{color,soul}
    \paperwidth=\dimexpr \paperwidth + 4cm\relax
    \oddsidemargin=\dimexpr\oddsidemargin + 2cm\relax
    \evensidemargin=\dimexpr\evensidemargin + 2cm\relax
    \marginparwidth=\dimexpr \marginparwidth + 2cm\relax
    \newcommand{\param}[1]{\textcolor{red}{#1}} % to highlight hardcoded numbers
    
    \newcommand{\agycomment}[1]{\todo[size=\scriptsize, linecolor=orange, bordercolor=orange, backgroundcolor=white]{\textcolor{gfored}{\textbf{@gy:} #1}}}

    \newcommand{\mscomment}[1]{\todo[size=\scriptsize, linecolor=orange, bordercolor=orange, backgroundcolor=white]{\textcolor{red}{\textbf{@MS:} #1}}}

    \newcommand{\atbcomment}[1]{\todo[size=\scriptsize, linecolor=orange, bordercolor=orange, backgroundcolor=white]{\textcolor{ao}{\textbf{@atb:} #1}}}

    \newcommand{\ouscomment}[1]{\todo[size=\scriptsize, linecolor=orange, bordercolor=orange, backgroundcolor=white]{\textcolor{ouscolor}{\textbf{@ous:} #1}}}

    \newcommand{\yctcomment}[1]{\todo[size=\scriptsize, linecolor=orange, bordercolor=orange, backgroundcolor=white]{\textcolor{yt}{\textbf{@yct:} #1}}}

    \newcommand{\gfcomment}[1]{\todo[size=\scriptsize, linecolor=orange, bordercolor=orange, backgroundcolor=white]{\textcolor{blue}{\textbf{@gf:} #1}}}

    \newcommand{\nbcomment}[1]{\todo[size=\scriptsize, linecolor=orange, bordercolor=orange, backgroundcolor=white]{\textcolor{nbs}{\textbf{@nb:} #1}}}

    \newcommand{\hluocomment}[1]{\todo[size=\scriptsize, linecolor=orange, bordercolor=orange, backgroundcolor=white]{\textcolor{moegi}{\textbf{@hluo:} #1}}}

    \newcommand{\iey}[1]{\textcolor{iy}{#1}}
    \newcommand{\ieyinline}[1]{{\color{iy}{\textbf{\hl{[@iey:}} \textit{\hl{#1}}\textbf{]}}}}
    \newcommand{\ieycomment}[1]{\todo[size=\scriptsize, linecolor=orange, bordercolor=orange, backgroundcolor=white]{\textcolor{iy}{\textbf{@iey:} #1}}}

    \newcommand{\omcomment}[1]{\todo[size=\scriptsize, linecolor=orange, bordercolor=orange, backgroundcolor=white]{\textcolor{teal}{\textbf{@ste:} #1}}}
    \newcommand{\ominline}[1]{{\textcolor{teal}{\textbf{[@ste:} #1\textbf{]}}}}

\else
    
    \newcommand{\param}[1]{\textcolor{black}{#1}} % to highlight hardcoded numbers
    \newcommand{\agycomment}[1]{}
    \newcommand{\agyinline}[1]{}

    \newcommand{\mscomment}[1]{}

    \newcommand{\atbcomment}[1]{}

    \newcommand{\yctcomment}[1]{}

    \newcommand{\gfcomment}[1]{}

    \newcommand{\nbcomment}[1]{}

    \newcommand{\ouscomment}[1]{}

    \newcommand{\hluocomment}[1]{}

    \newcommand{\iey}[1]{#1}
    \newcommand{\ieycomment}[1]{}
    \newcommand{\ieyinline}[1]{}

    \newcommand{\omcomment}[1]{}
    \newcommand{\ominline}[1]{}

\fi

\newcommand*\nCHIPS{96}
\newcommand*\nMODULES{12}

\newcommand*\obsv[1]{%
\refstepcounter{obs}
\noindent\textit{\underline{Observation~\theobs.}} \emph{#1}}

\newcounter{obs}
\newcounter{tkw}
\newcommand\takeawaybox[1]{%
    \vspace{1pt}
   \refstepcounter{tkw}
  \noindent
  \begin{tabular}{|p{0.95\linewidth}|}
       \hline
       \textbf{{Takeaway \thetkw}.} \emph{{#1}}\\
       \hline 
  \end{tabular}
  \addtocounter{table}{-1} % prevent incrementing table count
  \vspace{1pt}
}
\newcommand\vivekpud{\cite{seshadri2019dram, seshadri2017ambit, seshadri.bookchapter17, seshadri2013rowclone,seshadri2015fast,seshadri2016buddy, seshadri2016processing,seshadri2018rowclone}}

\newcommand{\figref}[1]{Fig.~\ref{#1}}

\newcommand{\secref}[1]{§\ref{#1}}

\newcommand*\circledtest[1]{\tikz[baseline=(char.base)]{
            \node[shape=circle,fill,inner sep=0.3pt] (char) {\textcolor{white}{#1}};}}

\ifcameraready

    \newcommand{\nbcr}[2]{\ifnum#1>-1\textcolor{black}{#2}\else{#2}\fi}
    \newcommand{\ieycr}[2]{\ifnum#1>-1\textcolor{black}{#2}\else{#2}\fi}
    \newcommand{\omcr}[2]{\ifnum#1>-1\textcolor{black}{#2}\else{#2}\fi}
    \newcommand{\omcrcomment}[1]{}
    \newcommand{\crdiscussion}[2]{}{}

    \newcommand{\ieycrcomment}[1]{}
    \newcommand{\atbcrcomment}[1]{}
    \newcommand{\agycrcomment}[1]{}
\else
\usepackage[colorinlistoftodos,prependcaption,textsize=scriptsize]{todonotes} % For margin notes
    \paperwidth=\dimexpr \paperwidth + 4cm\relax
    \oddsidemargin=\dimexpr\oddsidemargin + 2cm\relax
    \evensidemargin=\dimexpr\evensidemargin + 2cm\relax
    \marginparwidth=\dimexpr \marginparwidth + 2cm\relax

    \newcommand{\nbcr}[2]{\ifnum#1=\value{version}\textcolor{nbs}{#2}\else{#2}\fi}

    \newcommand{\ieycr}[2]{\ifnum#1=\value{version}\textcolor{blue}{#2}\else{#2}\fi}

    \newcommand{\ieycrcomment}[1]{\todo[linecolor=orange, bordercolor=orange, backgroundcolor=white]{\textcolor{iy}{\textbf{@Ismail:} #1}}}

    \newcommand{\atbcrcomment}[1]{\todo[linecolor=brown, bordercolor=brown, backgroundcolor=white]{\textcolor{ao}{\textbf{@Atb:} #1}}}

    \newcommand{\agycrcomment}[1]{\todo[size=\scriptsize, linecolor=orange, bordercolor=orange, backgroundcolor=white]{\textcolor{gfored}{\textbf{@gy:} #1}}}
    
    \newcommand{\crdiscussion}[2]{\omcrcomment{#1\\\textcolor{blue}{\textbf{@Ismail:}#2}}}
    \newcommand{\omcr}[2]{\ifnum#1=\value{version}\textcolor{red}{#2}\else{#2}\fi}

    \newcommand{\omcrcomment}[1]{\todo[linecolor=orange, bordercolor=orange, backgroundcolor=white]{\textcolor{red}{\textbf{@Onur:} #1}}}

\fi

\newcommand{\tras}[0]{t_{RAS}}
\newcommand{\trp}[0]{t_{RP}}
\newcommand{\trc}[0]{t_{RC}}
\newcommand{\trefi}[0]{t_{REFI}}
\newcommand{\trefw}[0]{t_{REFW}}
\newcommand{\trfc}[0]{t_{RFC}}
\newcommand{\trrd}[0]{t_{RRD}}

\newcommand{\sar}[0]{$SAR$}

\newcommand{\act}[0]{$ACT$}
\newcommand{\pre}[0]{$PRE$}
\newcommand{\refresh}[0]{$REF$}
\newcommand{\wri}[0]{$WR$}
\newcommand{\rd}[0]{$RD$}

\newcommand{\apatrng}[0]{$ACT$~$\rightarrow$~$PRE$~$\rightarrow$~$ACT$}

\newcommand{\apaLong}[0]{\act{} $\rightarrow$ \pre{} $\rightarrow$ \act{}}
\newcommand{\apa}[0]{$APA$}

\newcommand{\pum}[0]{\texttt{PuM}}

\usepackage[shortcuts]{extdash}

\newacronym{iqr}{$IQR$}{inter-quartile range}
\newacronym{simra}{SiMRA}{simultaneous multiple-row activation}
\newacronym{cots}{COTS}{commercial off-the-shelf}
\newacronym{act}{\act{}}{activate}
\newacronym{pre}{\pre{}}{precharge}
\newacronym{ref}{\refresh{}}{refresh}
\newacronym{wr}{\wri{}}{write}
\newacronym{rd}{\rd{}}{read}
\newacronym{pum}{\pum{}}{Processing-using-Memory}
\newacronym{apa}{\apa{}}{\act{} $\rightarrow$ \pre{} $\rightarrow$ \act{}}

\newacronym{jedec}{JEDEC}{Joint Electron Device Engineering Council}

\newacronym{trefw}{$\trefw$}{refresh window}
\newacronym{tras}{$\tras$}{XXXX}
\newacronym{trp}{$\trp$}{XXXX}
\newacronym{trc}{$\trc$}{the minimum time needed between two consecutive row activations targeting the same bank}
\newacronym{trefi}{$\trefi$}{refresh interval}
\newacronym{trfc}{$\trfc$}{refresh latency}
\newacronym{trrd}{$\trrd$}{the minimum time needed between two consecutive row activations targeting the same rank}
\newacronym{puf}{PUF}{physical unclonable function}
\newacronym{trn}{TRN}{true random number}

\begin{document}
\title{In-DRAM True Random Number Generation\\Using Simultaneous Multiple-Row Activation:\\An Experimental Study of Real DRAM Chips}
\newcommand{\affilETH}[0]{\textsuperscript{\S}}
\newcommand{\affilCISPA}[0]{\textsuperscript{$\Gamma$}}
\author{
{{\.I}smail Emir Y{\"u}ksel\affilETH}\qquad
{Ataberk Olgun\affilETH}\qquad
{F. Nisa Bostanc{\i}\affilETH}\qquad
{O\u{g}uzhan Canpolat\affilETH}\\
{Geraldo F. Oliveira\affilETH}\qquad
{Mohammad Sadrosadati\affilETH}\qquad
{A.~Giray~Ya\u{g}l{\i}k\c{c}{\i}\affilETH\affilCISPA}\qquad
{Onur Mutlu\affilETH}
\vspace{-2mm}\\\\
{\affilETH ETH Z{\"u}rich\qquad{} \affilCISPA CISPA}
}

\maketitle

    \renewcommand{\headrulewidth}{0pt}
    \fancypagestyle{firstpage}{
        \fancyhead{} 
        \fancyhead[C]{
      } 
    \renewcommand{\footrulewidth}{0pt}
    }
  \thispagestyle{firstpage}

\pagenumbering{arabic}

\newcounter{version}
\setcounter{version}{999}
\begin{abstract}

    In this work, we experimentally demonstrate that it is possible to generate true random numbers at high throughput and low latency in \gls{cots} DRAM chips by leveraging \gls{simra} \omcr{0}{via an} extensive characteriz\omcr{0}{ation of} \nCHIPS{} DDR4 DRAM chips\omcr{0}{. We} rigorously analyze \gls{simra}'s true random generation potential in terms of entropy, latency, and throughput for varying numbers of simultaneously activated DRAM rows (i.e., 2, 4, 8, 16, and 32), data patterns, temperature levels, and spatial variations. \ieycr{0}{Among our 11 key \omcr{0}{experimental} observations, w}e highlight \ieycr{0}{four} key results. 
    First, \ieycr{0}{we evaluate the quality of our TRNG designs using the commonly-used NIST statistical test suite for randomness and find that all \gls{simra}-based TRNG designs successfully pass each test. Second,} 2-, 8-, 16-, and 32-row activation-based TRNG designs outperform the state-of-the-art DRAM-based TRNG in throughput by up to \param{1.15}$\times$, \param{1.99}$\times$, \param{1.82}$\times$, and \param{1.39}$\times$, respectively.
    \ieycr{0}{Third}, \gls{simra}'s entropy tends to increase with the number of simultaneously activated DRAM rows. For example, for most of the tested modules, the average entropy of 32-row activation is \param{2.51$\times$}
    higher than that of 2-row activation.
    \ieycr{0}{Fourth}, operational parameters and conditions (e.g., data pattern and temperature) significantly affect entropy. For example, increasing temperature from 50$^{\circ}$C to 90$^{\circ}$C decreases SiMRA's entropy by \param{1.53$\times$} for 32-row activation. To aid future research and development, we open-source our infrastructure at \url{https://github.com/CMU-SAFARI/SiMRA-TRNG}.

\end{abstract}
\glsresetall
\section{Introduction}
True random number generators (TRNGs) harness entropy from random physical phenomena (e.g., electrical and thermal noise~\cite{gong2019true,huang2014real,srinivasan20102}) to generate unpredictable and irreproducible random bitstreams (i.e., true random numbers). True random numbers are crucial in many applications, including \ieycr{0}{cryptography, scientific simulations, machine learning, and recreational {entertainment}~\cite{darrell1998genetic, vincent2010stacked, schmidt1992feedforward, zhang2016survey, mich2010machine, bagini1999design, bakiri2018survey,
rock2005pseudorandom, ma2016quantum, stipvcevic2014true,
barangi2016straintronics, tao2017tvl, gutterman2006analysis, von2007dual,
kim2017nano, drutarovsky2007robust, kwok2006fpga, cherkaoui2013very,
zhang2017high, quintessence2015white, clarke2011robust}.}
These applications often require a high-throughput TRNG~\cite{wang2016theory, clarke2011robust, lu2015fpga,bostanci2022drstrange}. A dedicated hardware TRNG (i.e., an integrated circuit dedicated to true random number generation) can satisfy the throughput requirements of such applications.
Unfortunately, \emph{not} all computing systems have dedicated hardware TRNGs.

To alleviate \omcr{0}{the need for} dedicated TRNG hardware, using DRAM~\cite{dennard1968dram} as the entropy source for generating true random numbers (i.e., DRAM-based TRNG) is a promising approach due to the prevalence of DRAM throughout \omcr{0}{almost all} modern computing systems, ranging from microcontrollers to supercomputers\ieycr{0}{, and systems that employ Processing-in-Memory (PIM)~\cite{ghose.ibmjrd19, mutlu2020modern, mutlu2019processing, mutlu2024memory,mutlu2025memory,seshadri2017ambit,yuksel2024functionally,gomez2023evaluating,hajinazar2021simdram,deoliveira2024mimdram,li2017drisa,oliveira2025proteus,gomez2022benchmarking,devaux2019true,skhynixpim,kwon202125,niu2022184qps, ke2021near}.} Several prior works~\cite{keller2014dynamic,tehranipoor2016robust,eckert2017drng,kim2019d,talukder2019exploiting,pyo2009dram,olgun2021quactrng} demonstrate that by leveraging the analog operational properties of DRAM, it is possible to generate true random numbers in \gls{cots} DRAM chips.

Recent works~\cite{gao2019computedram,gao2022frac,olgun2021quactrng,yuksel2023pulsar,yuksel2024functionally,yuksel2024simultaneous} experimentally demonstrate
a new \ieycr{0}{phenomenon} in \gls{cots} DRAM chips that could potentially be used as a source of entropy:~\omcr{0}{\emph{\gls{simra}}}. 
\ieycr{0}{These works~\cite{gao2019computedram,gao2022frac,olgun2021quactrng,yuksel2023pulsar,yuksel2024functionally,yuksel2024simultaneous} carefully engineer
a sequence of DRAM commands that allows the DRAM chip
to activate (i.e., open) multiple (i.e., up to 32\omcr{1}{)} DRAM rows \omcr{1}{simultaneously}~\cite{yuksel2024simultaneous,yuksel2023pulsar})}, which is fundamentally different from the conventional single-row DRAM activation.

We hypothesize that \ieycr{0}{\gls{simra} could be used as the entropy source of DRAM-based TRNGs. This is because} manufacturing process variations could introduce high unpredictability in the charge-sharing process \ieycr{0}{when} multiple rows are activated simultaneously. Therefore, \gls{simra} could potentially provide high entropy and thereby be leveraged for high-throughput and low-latency true random number generation. To design efficient and high-throughput DRAM-based TRNGs, it is important to develop a better understanding of the entropy that \gls{simra} can provide.

In this work, our \emph{goal} is to experimentally study and understand the potential of \gls{simra} as a TRNG substrate in COTS DRAM chips.
To this end, we provide two main analyses in our study. 
First, we experimentally characterize the randomness (quantitatively measured using Shannon entropy~\cite{shannon1948mathematical}, described in \secref{subsec:testing_met}) that SiMRA provides in \nCHIPS{} COTS DDR4 DRAM chips \omcr{0}{while varying a large number of parameters}: the number of simultaneously activated rows, DRAM \ieycr{0}{chip density \& revision}, data pattern, temperature, and
spatial variation across DRAM subarrays. 
Second, we evaluate the quality \ieycr{0}{(using the standard NIST
statistical test suite for randomness~\cite{rukhin2001statistical})}, throughput, and latency of new TRNG designs that leverage \gls{simra} for varying numbers of simultaneously activated rows.

\ieycr{0}{Based on our real DRAM chip experiments, we make 11 new empirical
observations and share 5 key takeaway lessons.} We summarize \ieycr{0}{four} major results from our analyses.
\ieycr{0}{First, we observe that a TRNG that leverages \gls{simra} (SiMRA-TRNG) is a high-quality TRNG. We show that the random bitstreams generated by SiMRA-TRNG pass all the standard NIST statistical test suite randomness tests~\cite{rukhin2001statistical}.}
\ieycr{0}{Second}, we observe that SiMRA-TRNG generates true random numbers at high throughput and low latency. Our results show 
that \ieycr{0}{across all tested DRAM chips,} 2-, 8-, 16-, and 32-row activation-based SiMRA-TRNGs achieve up to \param{1.15}$\times$, \param{1.99}$\times$, \param{1.82}$\times$, and \param{1.39}$\times$ \ieycr{0}{(and on average \param{0.72}$\times$, \param{1.16}$\times$, \param{1.00}$\times$, and \param{0.78}$\times$)} higher throughput than the state-of-the-art \omcr{1}{DRAM-based TRNG}~\cite{olgun2021quactrng}, respectively. 
\ieycr{0}{Third}, the number of activated rows has a significant effect on entropy, and entropy increases with the number of activated rows in most of the tested DRAM chips. We observe that in \param{64} DRAM chips, \emph{increasing} the number of activated rows from 2 to 32 \emph{increases} the entropy by \param{2.51$\times$} on average. However, in the remaining \param{32} DRAM chips, we do \emph{not} observe a clear trend (i.e., the entropy can either increase or decrease with the number of activated rows).
\ieycr{0}{Fourth}, operating parameters and conditions (i.e., data pattern, temperature, and spatial variation) significantly affect the entropy of \gls{simra}.
For example, our results for 32-row activation show that increasing the temperature from 50$^{\circ}$C to 90$^{\circ}$C decreases the entropy on average across all tested DRAM chips by \param{1.53$\times$}. \omcr{0}{Our work is openly and freely available at \url{https://github.com/CMU-SAFARI/SiMRA-TRNG}.}

We make the following key contributions:
\begin{itemize}
    \item \omcr{0}{We provide the first experimental characterization of the ability of COTS DRAM chips to generate true random numbers via simultaneous multiple-row activation (\gls{simra}) for varying numbers of simultaneously activated rows.}
    \item We \ieycr{0}{rigorously} characterize \param{\nCHIPS{}} COTS DDR4 chips under various parameters and conditions\ieycr{0}{ \omcr{1}{in terms the entropy provided by \gls{simra}}: the number of simultaneously activated rows, DRAM chip density \& \omcr{1}{die} revision, data pattern, temperature, and
spatial variation across DRAM subarrays.}
    \item We demonstrate that 
    i)~SiMRA-TRNG generates high-quality and high throughput true random numbers; and
    ii)~2-, 8-, 16-, and 32-row activation-based SiMRA-TRNG improve\omcr{1}{s} throughput over the state-of-the-art \omcr{1}{DRAM-based TRNG}~\cite{olgun2021quactrng}.
\end{itemize}

\glsresetall
\section{Background}
We provide a brief background on DRAM, simultaneous multiple-row activation, and true random number generators. 

\subsection{Dynamic Random Access Memory (DRAM)}
\label{sec:dram_background}
\begin{figure}[ht]
\centering
\includegraphics[width=\linewidth]{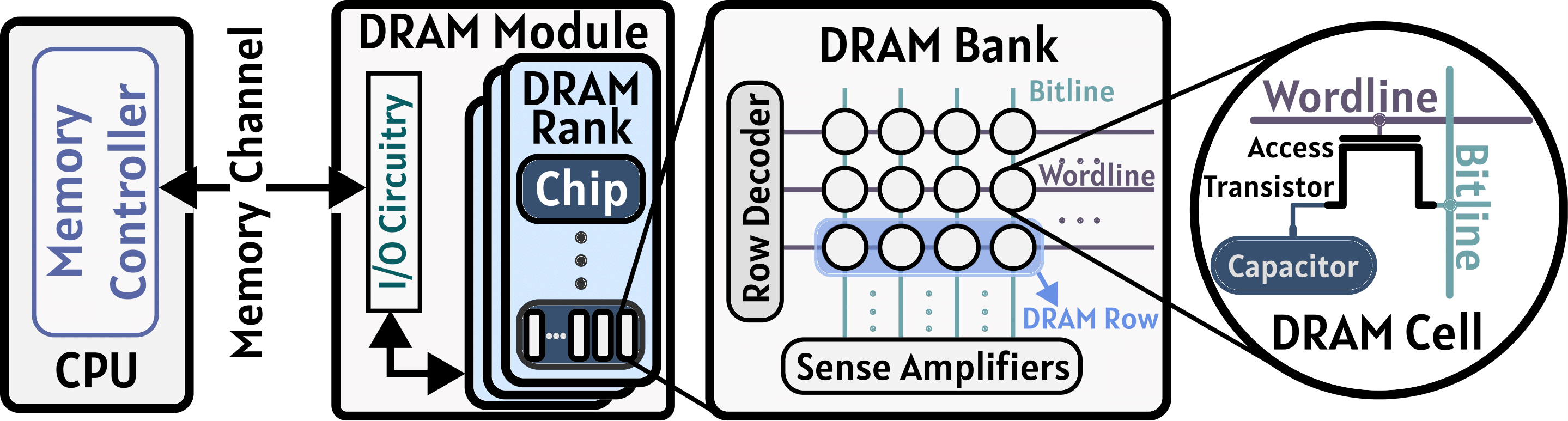}
\caption{DRAM organization. \omcr{0}{Adapted from~\cite{orosa2021deeper}.}}
\label{fig:dram_back}
\end{figure}

\noindent\textbf{DRAM Organization.} 
\figref{fig:dram_back} depicts the hierarchical organization of DRAM-based main memory. The memory controller in a system is connected to multiple DRAM modules via multiple DRAM channels. A DRAM module has one or more ranks, each of which consists of multiple DRAM chips that operate in lock-step. The memory controller can interface with multiple DRAM ranks by time-multiplexing the channel’s I/O bus between the ranks. Each chip has multiple DRAM banks, each consisting of multiple DRAM cell arrays (called subarrays\omcr{1}{~\cite{kim2012case, chang2014improving, seshadri2013rowclone, yaglikci2022hira}}). Within a subarray, DRAM cells are organized as a two-dimensional array of DRAM rows and columns. A DRAM cell stores one bit of data in the form of an electrical charge in a capacitor, which can be accessed through an access transistor. A wire called \emph{wordline} drives the gate\omcr{0}{s} of all DRAM cells’ access transistors in a DRAM row. A wire called \emph{bitline} connects all DRAM cells in a DRAM column to a common differential sense amplifier. Therefore, when a wordline is asserted, each DRAM cell in the
DRAM row is connected to its corresponding sense amplifier. The sense amplifier compares the voltage level of the activated cell's bitline and the reference voltage (VDD/2) and amplifies the voltage difference as logic-1 or logic-0. 

\noindent\textbf{DRAM \ieycr{0}{Commands}.}
To serve main memory requests \omcr{0}{and to maintain data integrity}, the memory controller issues DRAM commands, e.g., row activation (\act{}), bank precharge (\pre{}), data read (\rd{}), data write (\wri{}), and refresh (\refresh{}). 
To perform a read or write operation, the memory controller first needs to open a row, i.e., copy the data of the cells in the row to the \ieycr{0}{sense amplifiers}. 
To open a row, the memory controller issues an \act{} command to a bank by specifying the address of the row to open. After \omcr{0}{row} activation completes, the memory controller issues either a \rd{} or a \wri{} command to read or write a DRAM word within the activated row. 
To access data from another DRAM row in the same bank, the memory controller must first close the currently open row by issuing a \pre{} command. 
The memory controller also periodically issues \refresh{} commands to prevent data loss due to charge leakage.

\noindent\textbf{\ieycr{0}{DRAM Cell Operation \& Timing.}}
\ieycr{0}{We describe DRAM cell operations and DRAM timing by explaining the steps in activating a cell.\footnote{\omcr{1}{We describe DRAM cell operation, but all the described operations happen at row granularity (i.e., to all cells in the activated row).}} First, to activate the cell, the memory controller issues an ACT command. The row decoder asserts the wordline
and thus connects the cell capacitor to the bitline. Second, the cell
capacitor charge begins to perturb the bitline, known as the charge-sharing process. Charge sharing continues until the bitline voltages reach a level that the sense amplifier can safely amplify (e.g., VDD/2+$\epsilon$ or VDD/2-$\epsilon$). Third, the sense amplifier then kicks in to
amplify the difference between the bitline voltage and the reference voltage (VDD/2). Depending on the cell’s charge, the bitline becomes either VDD or GND. To return a cell to its precharged state (i.e., to restore the cell's charge), the voltage in the cell must first be fully restored, which requires waiting for $\tras{}$ timing parameter (i.e., the timing parameter between ACT and PRE commands). Fourth, once the cell is restored, the memory
controller can issue a PRE command. The cell returns to
the precharged state, where
the memory controller can reliably issue an ACT command after
waiting for timing parameter $\trp{}$ \omcr{1}{(i.e., the timing parameter between PRE and ACT commands)}.}

\subsection{Multiple-Row Activation in DRAM}

\ieycr{0}{Activating multiple DRAM rows (i.e., multiple-row activation) is a key technique to perform massively parallel in-DRAM computation, as initially shown in~\vivekpud{}. \omcr{1}{P}rior works~\cite{gao2019computedram,olgun2023dram,gao2022frac,yuksel2024simultaneous,olgun2022pidram,yuksel2024functionally, seshadri2013rowclone,seshadri2016buddy,seshadri2015fast,seshadri2016processing,seshadri2017ambit,seshadri2017simple,seshadri2018rowclone,seshadri2019dram,mutlu2024memory,mutlu2025memory,hajinazar2021simdram,deoliveira2024mimdram,oliveira2025proteus} \omcr{1}{perform} 1) in-DRAM data copy \& initialization (as in~\cite{seshadri2013rowclone}) and 2) in-DRAM bitwise operations (as in~\cite{seshadri2017ambit, seshadri2015fast}).} 

\noindent\textbf{\ieycr{0}{In-DRAM Data Copy \& Initialization.}}
\ieycr{0}{RowClone~\cite{seshadri2013rowclone} enables data movement within DRAM {at} row granularity without incurring the energy and execution time costs of transferring data between the DRAM and the computing units. {An intra-subarray} RowClone operation~\cite{seshadri2013rowclone} works by issuing {two} back-to-back \act{} commands: the first \act{} copies the contents of the source row $A$ into the {sense amplifiers}. The second \act{} connects the DRAM cells in the destination row~$B$ to the bitlines. Because the sense amplifiers have already sensed and amplified the source data by the time row~$B$ is activated, the data in each cell of row~$B$ is overwritten by the data stored in the {sense amplifiers} (i.e., row~$A$'s data).}

\noindent\textbf{\ieycr{0}{In-DRAM Bitwise Operations.}}
\ieycr{0}{Prior works~\cite{seshadri2017ambit,seshadri2015fast} introduce the concept of simultaneously activating three rows in a DRAM subarray (i.e., triple-row activation) through modifications to the DRAM circuitry. When three rows are concurrently activated, three cells connected to each bitline share charge simultaneously and contribute to the perturbation of the bitline~\cite{seshadri2017ambit,seshadri2015fast}. Upon sensing the perturbation of the three simultaneously activated rows, the sense amplifier amplifies the bitline voltage to $V_{DD}$ or 0~V if at least two of the three DRAM cells are charged or discharged, respectively. As such, simultaneously activating three rows results in a Boolean majority-of-three operation (MAJ3).}

\noindent\textbf{\ieycr{0}{SiMRA in COTS DRAM Chips.}} Many operations envisioned by these works~\cite{seshadri2017ambit,seshadri2015fast} can \emph{already} be performed in \emph{real unmodified} \gls{cots} DRAM chips. Recent works experimentally demonstrate that \gls{cots} DRAM chips can simultaneously activate multiple DRAM rows \ieycr{0}{(called SiMRA~\cite{yuksel2024simultaneous})} by issuing \omcr{0}{an} \apaLong{} (\apa{}) command sequence with violated $\tras{}$ and $\trp{}$ timing constraints~\cite{gao2019computedram,gao2022frac,olgun2021quactrng,olgun2022pidram,olgun2023dram,yuksel2023pulsar,yuksel2024functionally,yuksel2024simultaneous,yuksel2025pudhammer}. By doing so, \omcr{1}{\gls{cots}} DRAM chips can perform \one{} bitwise operations (i.e., MAJ, AND, OR, NOT, NAND, and NOR)~\cite{gao2019computedram,gao2022frac,yuksel2023pulsar,yuksel2024functionally,yuksel2024simultaneous, olgun2023dram}, \two{} data copy \& initialization~\cite{gao2019computedram, olgun2022pidram, yuksel2023pulsar, yuksel2024simultaneous}, and \three{} true random number generation~\cite{olgun2021quactrng,olgun2022pidram}.

\subsection{True Random Number Generators \omcr{0}{(TRNGs)}}
\label{subsec:trng}
A true random number generator (TRNG) samples random physical processes (e.g., thermal noise) to construct a random bitstream.
Harvesting randomness from a physical phenomenon \emph{often} produces bits that are biased or correlated~\cite{kocc2009cryptographic, rahman2014ti}. Post-processing methods can eliminate the bias and correlation and provide
protection against environmental changes and adversary tampering~\cite{kocc2009cryptographic, rahman2014ti, stipvcevic2014true}. Well-known post-processing techniques are the von Neumann corrector~\cite{von195113} and cryptographic hash functions such as SHA-256~\cite{fips2012180}.

\noindent\textbf{\ieycr{0}{DRAM-based TRNGs.}}
\ieycr{0}{Prior works propose DRAM-based mechanisms to implement true random number generators (TRNGs) by violating timing parameters~\cite{kim2019d,talukder2019exploiting,olgun2021quactrng}, using {data} retention failures~\cite{keller2014dynamic,sutar2018d}, and using startup values~\cite{eckert2017drng,tehranipoor2016robust}. The state-of-the-art DRAM-based TRNG~\cite{olgun2021quactrng} generates true random numbers by simultaneously activating four rows (called QUAC). }
\section{Experimental Methodology}

\subsection{COTS DRAM Testing Infrastructure} 
\begin{figure}[ht]
    \centering
    \includegraphics[width=1\linewidth]{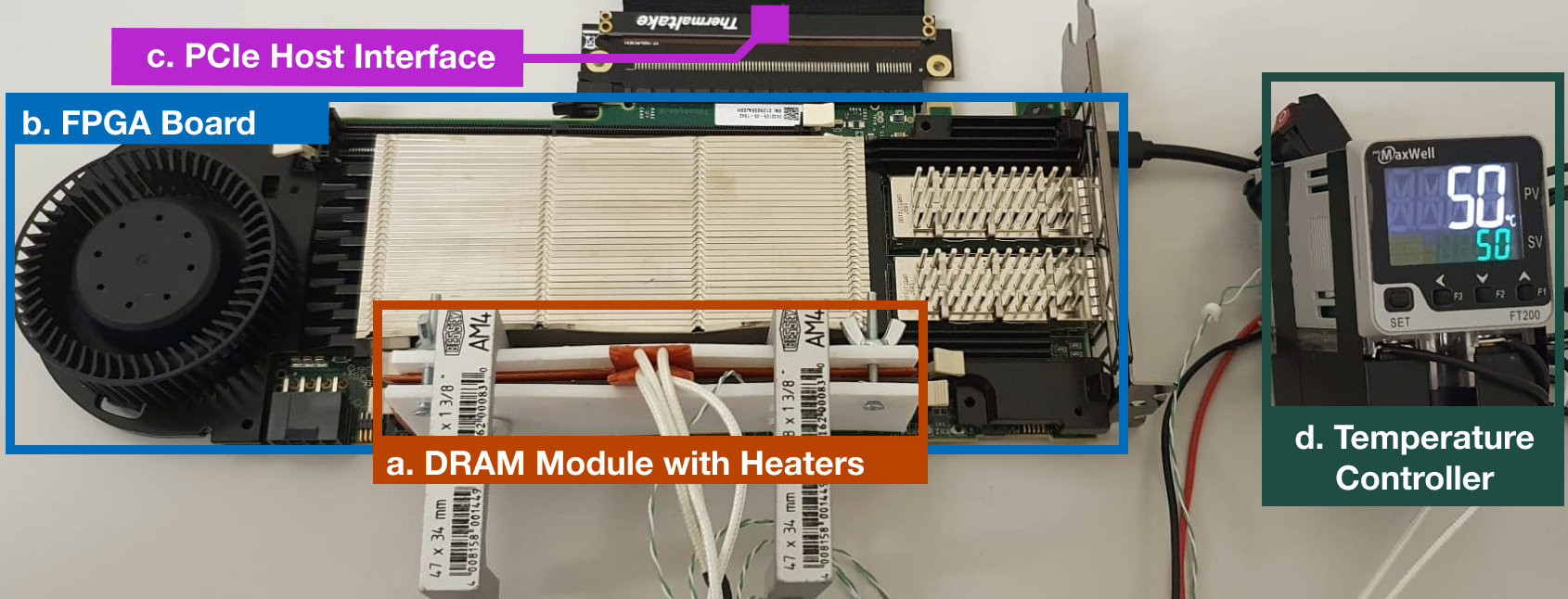}
    \caption{Our DRAM Bender~\cite{olgun2023dram} based experimental setup.}
    \label{fig:infra}
\end{figure}

\begin{figure}[ht]
\centering
\includegraphics[width=\linewidth]{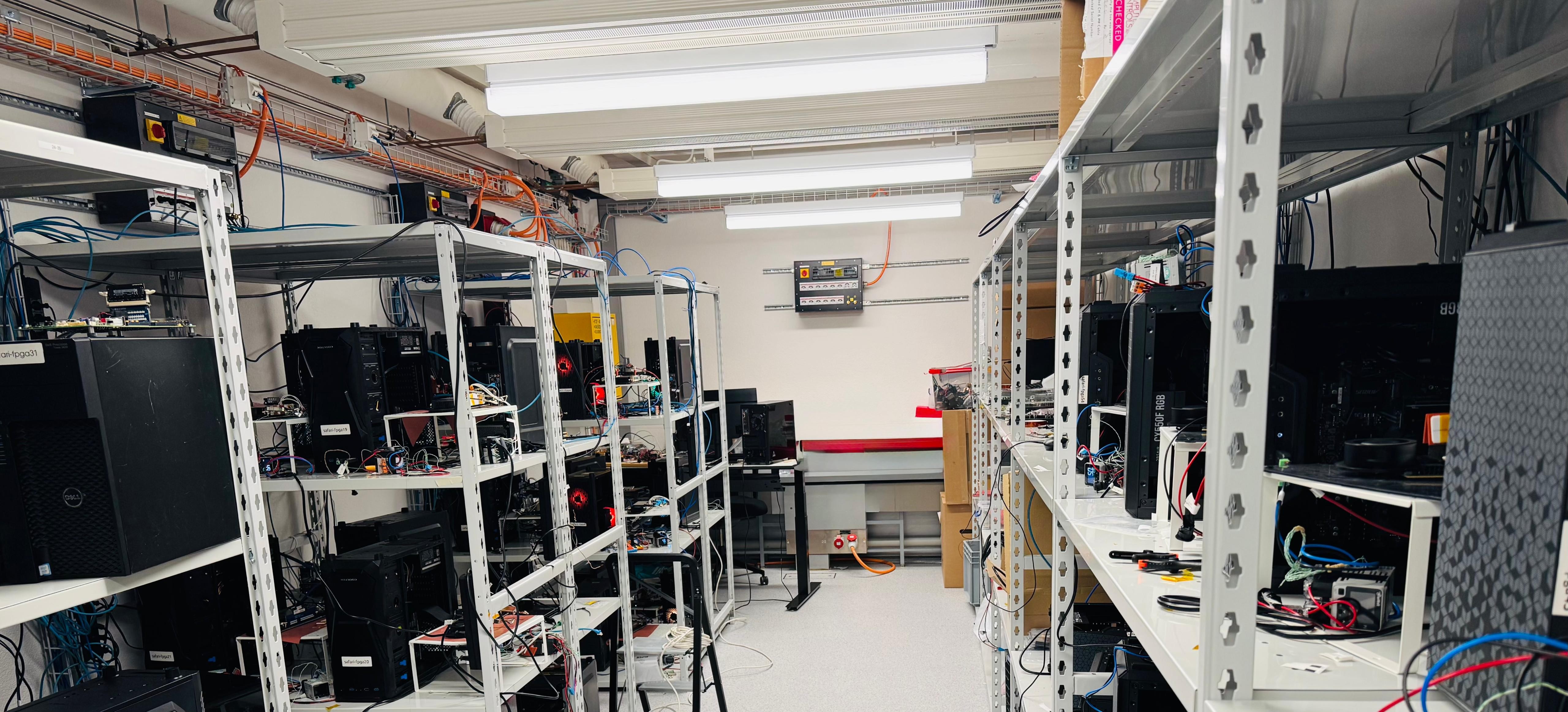}
\caption{\ieycr{0}{Our laboratory {for} real DRAM chip experiments.}}
\label{fig:lab}
\end{figure}

We conduct commercial off-the-shelf (COTS) DRAM chip experiments using DRAM Bender~\cite{safari-drambender, olgun2023dram} (built upon SoftMC~\cite{hassan2017softmc,softmcgithub}), an FPGA-based DDR4 DRAM testing infrastructure that provides precise control of the DRAM commands issued to a DDR4 DRAM module and DRAM timing parameters.~\figref{fig:infra} shows our experimental setup that consists of four main components:
a)~a thermocouple-based temperature sensor and a pair of heater pads pressed against the DRAM chips that heat up the DRAM chips to a desired temperature, 
b)~an FPGA development board (Xilinx Alveo U200~\cite{alveo}, programmed with DRAM Bender to execute our test programs,
c)~a host machine that generates the test program and collects experimental results,
and d)~a PID temperature controller (MaxWell FT200~\cite{maxwellFT200}) that controls the heaters and keeps the temperature at the desired level. \ieycr{0}{\figref{fig:lab} shows our
laboratory comprising many DDR4 \omcr{1}{and} HBM2 testing platforms.}

\noindent\textbf{COTS DDR4 DRAM Chips Tested.} Table~\ref{tab:dram_chips} provides the tested \param{\nCHIPS{}} (\param{\nMODULES{}}) COTS DDR4 DRAM chips (modules) that we focus our analysis on along with the chip manufacturer (Chip Mfr.), module manufacturer (Module Mfr.), the number of tested modules (\#Modules), the number of tested chips (\#Chips), die revision (Die Rev.), chip density, chip organization (Chip Org.), and DRAM module’s manufacturing date in the form of year-week (Date).

To investigate whether our characterization study applies to different DRAM technologies, designs, and manufacturing processes, we test a total of 152 (22) COTS DDR4 DRAM chips (modules) from all three major manufacturers (i.e., SK Hynix, Samsung, and Micron) with different die densities and die revisions from each DRAM chip manufacturer. 
While we observe successful \omcr{1}{\gls{simra}-based} random number generation in \omcr{0}{\emph{all}} tested SK Hynix modules, we observe no successful \omcr{0}{simultaneous} multiple-row activation (and thus random number generation)
in Samsung and Micron chips. \ieycr{0}{As \omcr{1}{described} in~\cite{yuksel2024simultaneous}, the reason we do not observe successful SiMRA operation may be that some DRAM chips contain internal circuitry that either ignores the \pre{} command or suppresses the second \act{} command when the timing parameters ($\trp{}$ and $\tras{}$) are significantly violated.}
Thus, we focus our analysis on \param{\nCHIPS{}} (\param{\nMODULES{}}) COTS DDR4 DRAM chips (modules) from SK Hynix, listed in Table 1.

\begin{table}[h!]
\resizebox{\linewidth}{!}{%
\centering
\begin{threeparttable}
\caption{Summary of DDR4 DRAM chips tested.}
\label{tab:dram_chips}
\begin{tabular}{@{}ccccccc@{}}
\toprule
\multirow{2}{*} {{\bf Chip Mfr.}} & {{\bf Module}} & \textbf{\#Modules} & {{\bf Die}}  &{{\bf Chip}} & {{\bf Chip}} & {{\bf Date\tnote{$\nabla$}}} \\ 
& {{\bf Mfr.}} & \textbf{(\#Chips)} & {{\bf Rev.}}  & {{\bf Density}} & {{\bf Org.}} & {{\bf year-week}} \\ 
\midrule
\midrule
\multirow{3}{*}{SK Hynix} & TimeTec     & 4 (32) & A  & 4Gb & x8 & N/A \\ 
                          & TeamGroup   & 4 (32) & M  & 4Gb & x8 & N/A \\
                          & SK Hynix    & 4 (32) & A  & 8Gb & x8 & 18-43 \\ 
\bottomrule
\end{tabular}%
%}
\begin{tablenotes}
\item[$\nabla$] We report “N/A” if no date is marked on the label of a module.
\end{tablenotes}
\end{threeparttable}
}
\end{table}

\subsection{Testing Methodology}
\label{subsec:testing_met}
\noindent\textbf{DRAM Subarray Boundaries.} 
Understanding which rows are simultaneously activated in the same subarray requires reverse engineering DRAM subarray boundaries. We follow the methodology used in prior works to identify subarray boundaries~\cite{gao2019computedram,olgun2022pidram,yuksel2024functionally,yuksel2024simultaneous,yuksel2023pulsar,olgun2021quactrng,yaglikci2022hira,yaglikci2024svard,mutlu2024memory,mutlu2025memory}.
We leverage the observation that COTS DRAM chips can copy the content of a DRAM row (i.e., source row) to another DRAM row (i.e., destination row) if \omcr{0}{and only if} these rows are in the \omcr{0}{\emph{same}} subarray. We perform the in-DRAM copy operation for every possible source and destination row address in each tested bank and reverse engineer the DRAM subarray boundaries.

\noindent\textbf{Simultaneously Activated Rows with APA.} 
Prior works~\cite{olgun2021quactrng, yuksel2024functionally, yuksel2024simultaneous,yuksel2025pudhammer} \ieycr{0}{demonstrate that issuing an \apaLong{} (\apa{}) sequence allows DRAM chips to activate 2, 4, 8, 16, and 32 DRAM rows simultaneously. To uncover which rows are simultaneously activated by issuing an \apa{} sequence, these works~\cite{olgun2021quactrng, yuksel2024functionally, yuksel2024simultaneous,yuksel2025pudhammer}} show that issuing an \apa{} sequence \omcr{0}{followed by} a WRITE (WR) command overwrites the simultaneously activated rows with the data sent with the WR command. We follow the same methodology and reverse engineer the simultaneously activated rows with \apa{} sequence for every possible row address \ieycr{0}{in all tested subarrays}. We call a group of simultaneously activated rows as a simultaneously activated row (\sar{}) group.

\begin{figure}[ht]
\centering
\includegraphics[width=\linewidth]{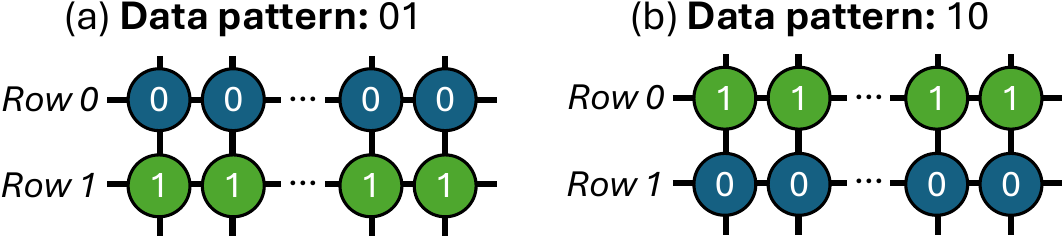}
\caption{\ieycr{0}{Two example data layouts for 2-row activation with (a) data pattern \texttt{01} and (b) data pattern \texttt{10}.}}
\label{fig:pattern}
\end{figure}

\noindent\textbf{Characterization Methodology.} 
Our characterization experiment consists of \param{three} key steps. First, we initialize rows in the tested \sar{} group with a predefined data pattern. Each bit in the data pattern is used to initialize one row in the \sar{} group. \nbcr{0}{\figref{fig:pattern} shows two example data layouts for 2-row activation with data patterns \texttt{01} (a) and \texttt{10} (b). We initialize the first \ieycr{0}{activated row with the left most bit value (e.g., for \texttt{01}, all cells in the first row have logic-0)} and second row with the right most bit value.} Second, we issue the \apa{} command sequence with reduced timings to simultaneously activate these rows. Third, we read the data in the tested subarray's sense amplifiers (a total of 65536 sense amplifiers) by respecting the timing parameters. We repeat this experiment for each \sar{} group for 1000 times and record the results in the host machine. 

\noindent\textbf{Entropy Metric.} We measure the randomness in DRAM sense amplifiers using Shannon entropy~\cite{shannon1948mathematical}. 
Shannon entropy ($H(x)$) of a bitstream generated by sampling a random source, which is sense amplifier ($x$) in our experiments, is calculated as $H(x) = -\sum_{i = 0}^{1}P(x_{i})\log_{2}P(x_{i})$, where $P(x_{0})$ and $P(x_{1})$ represent the probabilities of observing logic-0 and logic-1, respectively. We measure the Shannon Entropy by repeatedly performing the characterization experiment 1000 times for a tested \sar{} group. 
We analyze the entropy \omcr{0}{at the} granularity of 512-bit blocks (also known as cache blocks), as the \rd{} and \wri{} operation granularity \omcr{0}{of a DRAM module attached to a CPU is at cache block (e.g., 64B) granularity}. To do so, we calculate the entropy of each cache block by aggregating the entropy values of the bitstream generated by all 512 sense amplifiers corresponding to that cache block.
We define a metric called \emph{average cache block entropy}: the average entropy across all cache blocks in a row of sense amplifiers (i.e., 128 cache blocks) in a \sar{} group. For example, \omcr{0}{for} a \sar{} group with two cache blocks having entropies of 25 and 75 \omcr{0}{each}, the average cache block entropy group is 50.

\noindent\textbf{Number of Tested SAR Groups.} 
We observe that a tested DRAM bank has between 64 and 128 subarrays, depending on the DRAM module. 
We randomly select three subarrays in one bank per DRAM module. Within each subarray, we randomly test 100 \sar{} groups each for 2-, 4-, 8-, 16-, and 32-row activation. 

\noindent\textbf{Data Pattern.} We analyze entropy
using data patterns with a varying number of logic-1 bits. \nbcr{0}{We sweep all possible numbers of logic-1 bits in the data pattern (i.e., from 0 to the number of rows that are activated). For each number of logic-1s, we first calculate the number of distinct data patterns. For example, 2-row activation has 2 distinct data patterns (i.e., \texttt{01} and \texttt{10}). If the number of data patterns is smaller than 100, we test all possible data patterns. Otherwise, we generate and test 100 random distinct data patterns to limit the larger parameter space  (e.g., 32-row activation has more than 4.2B distinct data patterns).}

\noindent\textbf{Temperature.} We perform our experiments at five temperature levels: 50$^{\circ}$C, 60$^{\circ}$C, 70$^{\circ}$C, 80$^{\circ}$C, and 90$^{\circ}$C. All experiments are conducted at 50$^{\circ}$C unless stated otherwise.

\ieycr{0}{To aid future
research and development, we open-source our infrastructure \omcr{1}{and scripts} at \url{https://github.com/CMU-SAFARI/SiMRA-TRNG}.}

\section{Using Simultaneous Multiple-Row\\Activation as a Source of Entropy}

\subsection{Key Mechanism}

\noindent\textbf{\ieycr{0}{Simultaneously Activating Up to 32 DRAM Rows.}}
\ieycr{0}{Our key idea relies on a fundamental design {characteristic} in modern DRAM chips: {hierarchical design of the row decoder circuitry}. Modern DRAM chips have multiple {levels} of row address decoding stages to reduce latency, area, and power consumption~\cite{bai2022low,weste2015cmos,turi2008high, olgun2021quactrng,dram-circuit-design,yuksel2024simultaneous}. T{he hierarchical row decoder structure} expands a row address into a set of intermediate control signals and issuing an \act{} command asserts multiple control signals~\cite{bai2022low,weste2015cmos, olgun2021quactrng,yuksel2024simultaneous,dram-circuit-design}. Prior works~\cite{yuksel2023pulsar, yuksel2024simultaneous} demonstrate that by changing the row addresses targeted by two \act{} commands in \omcr{1}{\apaLong{} (}\apa{}), we can control the number and addresses of the
simultaneously activated rows in a subarray. These works~\cite{yuksel2023pulsar, yuksel2024simultaneous} hypothesize that to simultaneously activate $2^N$ rows, N decoders in the hierarchical row decoder need to have asserted immediate signals (e.g., asserting intermediate signals in two decoders activates $2^2=4$ rows simultaneously). In our tested DRAM chips, the hierarchical row decoder circuitry consists of 5 decoders (the same as prior works~\cite{yuksel2023pulsar, yuksel2024simultaneous}), and thus, we can simultaneously activate up to $2^5=32$ rows. Due to space limitations, we refer the reader to these prior works~\cite{yuksel2023pulsar, yuksel2024simultaneous} for
more details on the hypothetical row decoder circuitry in DRAM
chips and how this row decoder design allows simultaneous activation of multiple rows.}

\begin{figure}[ht]
    \centering
    \includegraphics[width=1\linewidth]{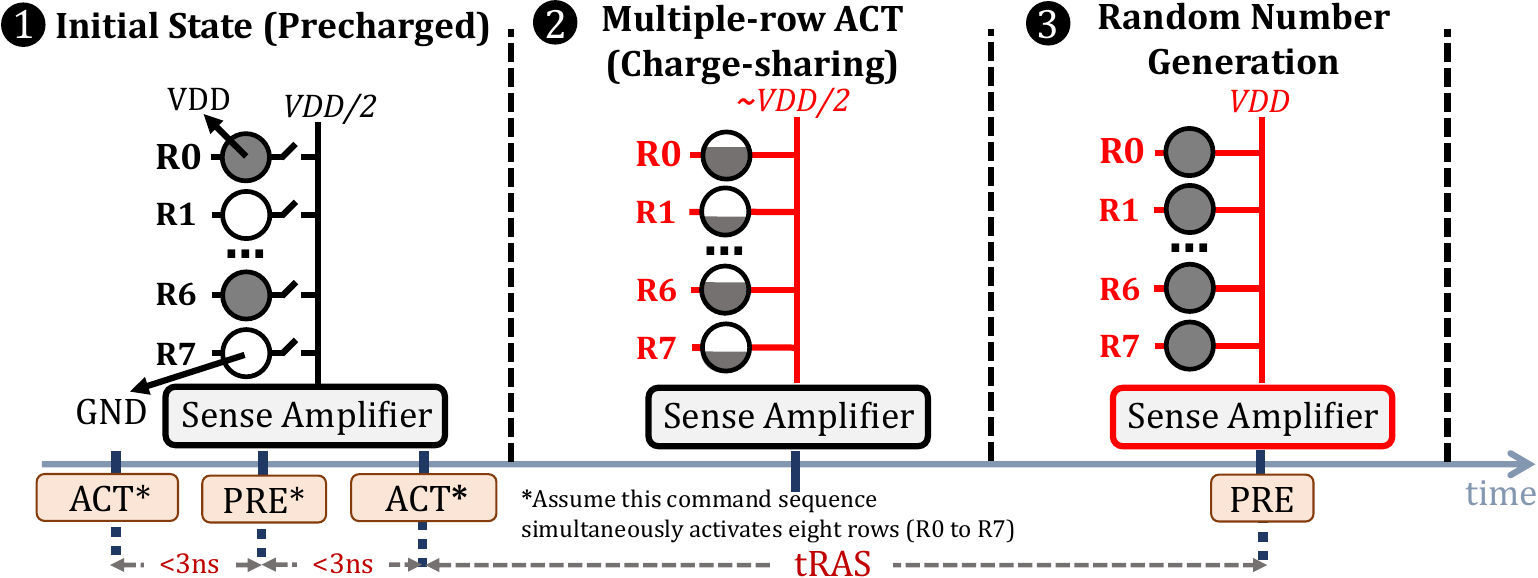}
    \caption{Command sequence for true random number generation in COTS DRAM chips using \gls{simra} and the state of cells during each related step.}
    \label{fig:mra_trng_mech}
\end{figure}

\noindent\textbf{\ieycr{0}{Key Idea.}}
\figref{fig:mra_trng_mech} shows the command sequence for true random number generation in COTS DRAM chips using \gls{simra} and eight DRAM cells connected to a bitline. The memory controller issues each command (shown in orange boxes below the time axis) at the corresponding tick mark, and asserted signals are highlighted in red.
Initially, four cells (R0, R2, R4, and R6) have a voltage level of VDD, and the remaining four cells (R1, R3, R5, and R7) ground (GND), and the bitline has a voltage level of VDD/2 (\dingOne{}). \ieycr{0}{To ease understanding, in \figref{fig:mra_trng_mech}, \omcr{1}{we} assume that \apatrng{} simultaneously activates eight rows (i.e., all rows from R0 to R7).}

To generate random numbers, we first issue one \apatrng{} command sequence (\dingOne{}). Doing so activates eight cells simultaneously and enables charge-sharing between \ieycr{0}{them and} their bitline. 
Hence, \ieycr{0}{the bitline voltage perturbation is affected by the simultaneously activated eight cells, where four of them try to pull up the bitline \omcr{1}{voltage} and the remaining four try to pull down the bitline voltage. As a result, these opposing contributions of simultaneously activated cells can randomly perturb the bitline \omcr{1}{from the reference voltage} $\sim$VDD/2 (\dingTwo{}).}
The sense amplifier then kicks in and tries to amplify the voltage on the bitline, which results in sampling a random value based on random perturbations on the bitline voltage~\cite{olgun2021quactrng, kim2019d}. For example, the single depicted bitline is randomly sampled as VDD in this figure (\dingThree{}). Finally, we send a \pre{} command to complete the process.

\subsection{COTS DRAM Chip Characterization Results}

\begin{figure}[ht]
\centering
\includegraphics[width=\linewidth]{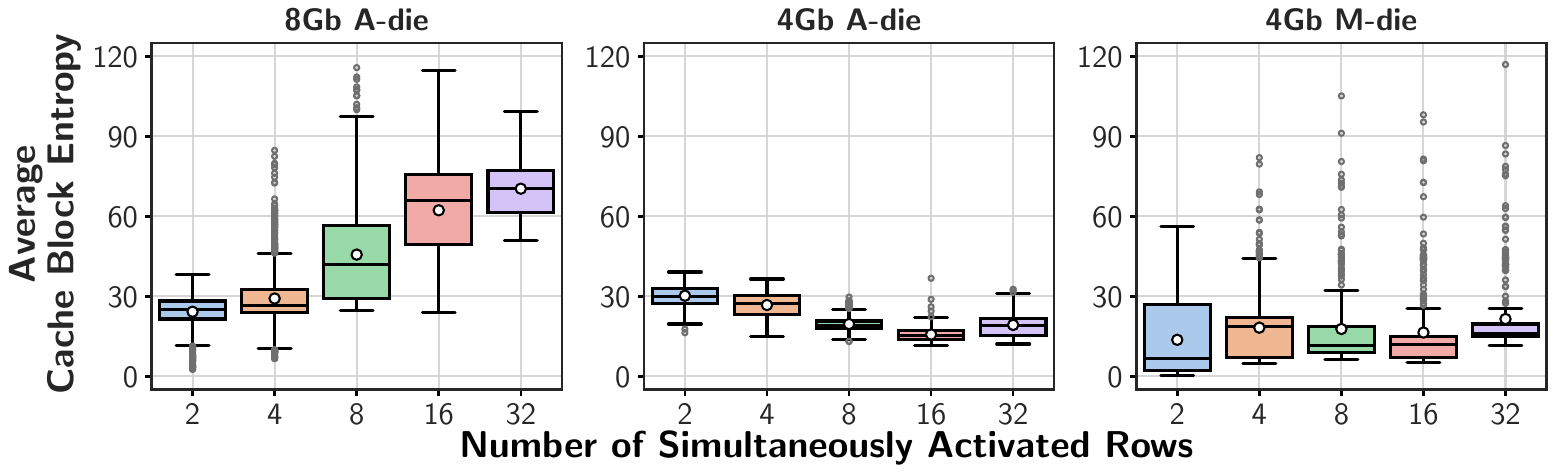}
\caption{The effect of DRAM architecture (chip density \& revision) and the number of simultaneously activated rows on the distribution of average cache block (512-bit) entropy.}
\label{fig:timing}
\end{figure}

\noindent\textbf{Effect of the Number of Simultaneously Activated Rows.} 
\figref{fig:timing} shows the distribution of the average cache block entropy (y-axis) for varying numbers of simultaneously activated rows in a box and whiskers plot.\footnote{\label{fn:boxplot}{In a box-and-whiskers plot, the box is lower-bounded by the first quartile and upper-bounded by the third quartile. The box size is the distance between the first and third quartiles. Whiskers show the minimum and maximum values.}} Each subfigure in \figref{fig:timing} represents \omcr{0}{a} different chip density and die revision pair. We make \textit{Observations 1-3} from \figref{fig:timing}. 

\obsv{\gls{simra} can generate random values with all tested numbers of simultaneously activated rows.} 

We observe that across all tested chips, \gls{simra} can generate \param{22.87}, \param{25.17}, \param{29.47}, \param{34.54}, and \param{40.41} average cache block entropy with 2-, 4-, 8-, 16-, and 32-row activations, respectively.

\obsv{Entropy tends to increase with the number of simultaneously activated rows.}

We observe that in 8Gb A-die and 4Gb M-die chips, the average cache block entropy increases by \param{2.51}$\times$ on average when we increase the number of simultaneously activated rows from 2 to 32. However, in 4Gb A-die chips (middle subplot), there is no strong correlation between the number of activated rows and cache block entropy as the entropy initially decreases until the number of rows reaches 16 and then increases when moving from 16- to 32-row activation.

We hypothesize that this could be due to two reasons. 
First, as it is impractical to test all possible data patterns for a higher number of row activations (e.g., \omcr{0}{16- or} 32-row activation), it is possible that we could not find the best data pattern to achieve the highest entropy. However, for a lower number of row activations (e.g., 2-row activation), \ieycr{0}{we explore} nearly all data patterns \ieycr{0}{since it} is more feasible \ieycr{0}{to explore all data patterns}, and thus, it is more likely \omcr{0}{that we} find the best data pattern to achieve the highest entropy. 
Second, many components in the DRAM circuitry could vary for each chip density \& \omcr{0}{die} revision pair, resulting in a different interaction between entropy and the number of simultaneously activated rows. For example, the row decoder circuitry may not fully drive the wordlines when more rows are activated, or the sense amplifier's safety margin could be wider in one DRAM architecture compared to others. We call for future circuit-level \omcr{0}{and device-level} research to fundamentally understand and gain more insights into how different DRAM components could affect \gls{simra}'s entropy.

\obsv{DRAM architecture (especially chip density) significantly affects \gls{simra}'s entropy.}

We observe that while 8Gb A-die \omcr{0}{has an} average cache block entropy of \param{37.62} across all numbers of activated rows, 4Gb A-die and M-die 
\omcr{0}{have} \param{25.06} and \param{16.83}, respectively.
We hypothesize this could be due to two reasons. First, newer (and denser) chips are more susceptible to \omcr{0}{read}
disturbance~\cite{orosa2021deeper,kim2020revisiting,kim2014flipping, yaglikci2022understanding, lim2017active, park2016statistical, park2016experiments, ryu2017overcoming, yun2018study, lim2018study, luo2023rowpress, lang2023blaster, yaglikci2024svard, nam2024dramscope, olgun2023hbm,nam2023xray,tugrul2025understanding,yuksel2025pudhammer,he2023whistleblower,luo2024experimental,olgun2025variable, girayphd,luo2025revisiting,yuksel2025columndisturb}, which could increase the noise on bitline voltage and result in increased entropy, compared to chips with lower density.
Second, as discussed in \emph{Observation~\param{2}}, \ieycr{0}{the significant entropy variation observed across DRAM chip densities could be due to differences in their underlying DRAM \omcr{1}{micro}architectures (i.e., DRAM chips of different densities could employ different DRAM \omcr{1}{micro}architectures).}

\takeawaybox{\gls{simra}'s entropy generally increases as the number of simultaneously activated rows increases\omcr{0}{. Entropy} and its interaction \omcr{0}{with} the number of simultaneously activated rows varies across DRAM \omcr{1}{micro}architectures.}

\begin{figure}[ht]
\centering
\includegraphics[width=\linewidth]{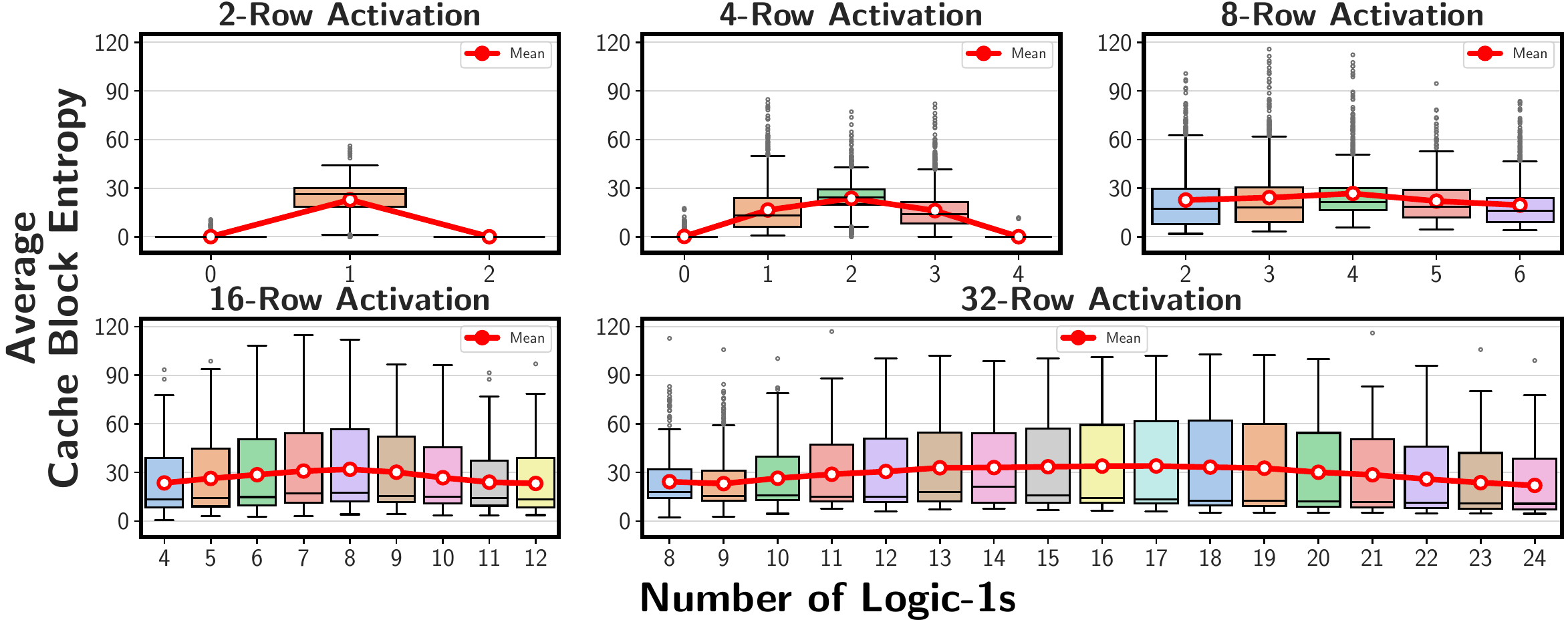}
\caption{Effect of the number of logic-1s in the data pattern on the distribution of average cache block entropy.}
\label{fig:numones}
\end{figure}

\noindent\textbf{Data Pattern Dependence.} \figref{fig:numones} shows the distribution of average cache block entropy (y-axis) across all tested DRAM chips for different numbers of simultaneously activated rows (subplots) with tested numbers of logic-1s in the data pattern (x-axis). We make \textit{Observation 4} from \figref{fig:numones}.

\obsv{The average cache block entropy significantly varies across tested numbers of logic-1s for every number of row activations.}

We observe that 2-, 4-, 8-, and 16-row activations produce the highest average cache block entropy when the data pattern has an equal number of logic-1s and logic-0s, which are \param{446.19}$\times$, \param{1.44}$\times$, \param{1.10}$\times$, and \param{1.03}$\times$ higher than the second highest average cache block entropy point, respectively. That is, the closer a data pattern is to the middle on the x-axis, the higher the entropy.
For 32-row activation, the highest average cache block entropy is produced at data pattern with 17 logic-1s, \param{1.01}$\times$ higher than the data patterns with 16 logic-1s. We hypothesize that data patterns that contain equal numbers of logic-1s and logic-0s (or one more logic-1) result in a small differential voltage. During charge sharing, logic-1 \omcr{0}{cells} attempt to raise the bitline voltage, while logic-0 \omcr{0}{cells} attempt to lower it.
Thus, the bitline voltage converges to the reference voltage, and the differential voltage is well below the sense amplifier's reliable sensing margin. Consequently, the sense amplifier fails to reliably \omcr{0}{sense \&} amplify, and \omcr{0}{thus,} non-deterministically settles to logic-1 or logic-0. 

\takeawaybox{Data pattern significantly affects \gls{simra}'s average cache block entropy.}

\noindent\textbf{Temperature.} 
In this experiment, we use a single \sar{} group with the configuration 
(defined by the combination of test parameters used, e.g., data pattern)
that achieves the highest cache block entropy for varying numbers of row activations in three tested modules, each from a different chip density \& die revision. \ieycr{0}{We select each module that achieves the highest average cache entropy among all chips of the same chip density \& \omcr{1}{die} revision and} test each \sar{} group 100 times.

\begin{figure}[ht]
\centering
\includegraphics[width=\linewidth]{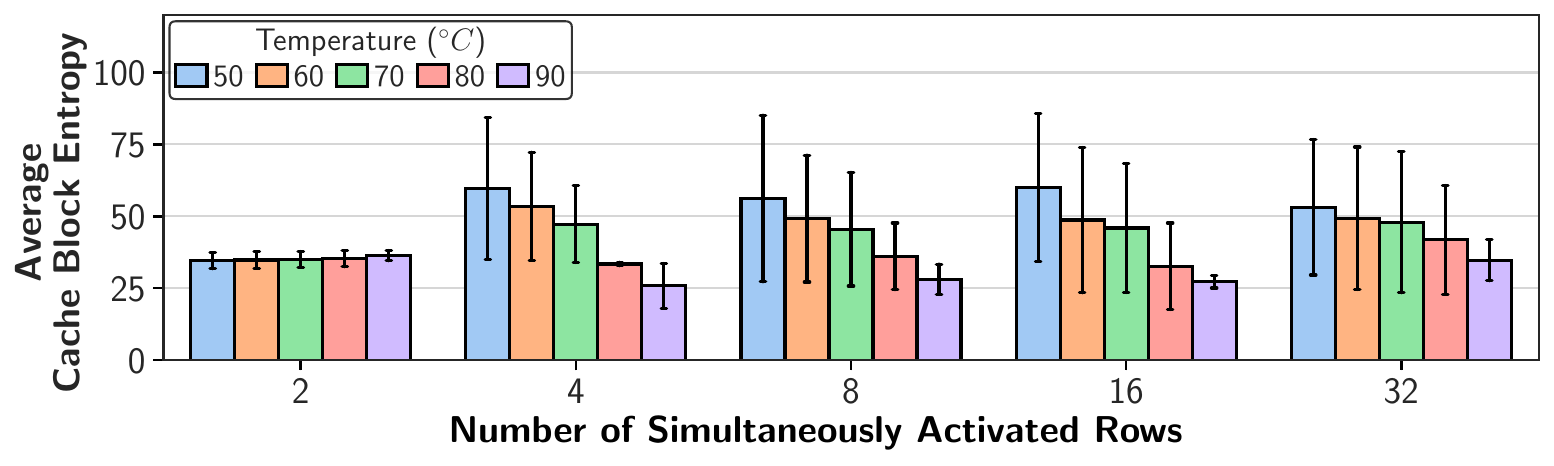}
\caption{Average cache block entropy at five temperature levels \omcr{0}{as a function of number of simultaneously activated rows}.}
\label{fig:temperature}
\end{figure}

\figref{fig:temperature} shows the average cache block entropy, with error bars representing the range across all tested DRAM modules for five DRAM chip temperature levels: 50$^{\circ}$C, 60$^{\circ}$C, 70$^{\circ}$C, 80$^{\circ}$C, and 90$^{\circ}$C. The figure clusters boxes into five groups based on the number of simultaneously activated rows (x-axis). In each cluster of boxes, temperature increases from left to right. We make \textit{Observation 5} from \figref{fig:temperature}.

\obsv{Entropy decreases as temperature increases for almost every number of row activations.}

For example, for 32-row activation, as the temperature increases from 50$^{\circ}$C to 90$^{\circ}$C, average cache block entropy decreases by \param{1.53$\times{}$}. 
We hypothesize that a high temperature could reduce the threshold voltage of access transistors and, thus, change the behavior of \omcr{0}{the} charge-sharing operation (e.g., by making cells share their charge more and faster)~\cite{yuksel2024simultaneous,sakata1994subthreshold,kao2002subthreshold}.
As a result, the best parameters (e.g., data pattern, timing parameters) to achieve the highest entropy could vary for different temperature levels.

\takeawaybox{DRAM chip temperature significantly affects \gls{simra}'s average cache block entropy.}

\noindent\textbf{Spatial Variation Across DRAM Subarrays.} 
In this experiment, we categorize each {tested subarray} into three groups based on \omcr{0}{its} location in a {bank}: \emph{Beginning} (i.e., the first 1/3 {subarrays of a bank}), \emph{Middle} (i.e., the second 1/3 {subarrays of a bank}), and \emph{End} (i.e., the last 1/3 {subarrays of a bank}).\footnote{\ieycr{0}{DRAM manufacturers use mapping schemes to translate logical (memory-controller-visible) addresses to physical row addresses~\cite{kim2014flipping, smith1981laser, horiguchi1997redundancy, keeth2001dram, itoh2013vlsi, liu2013experimental,seshadri2015gather, khan2016parbor, khan2017detecting, lee2017design, tatar2018defeating, barenghi2018software, cojocar2020rowhammer,  patel2020beer,kim2012case,frigo2020trrespass}. To account for in-DRAM row address mapping, we reverse engineer the physical row address layout {in all chips}, following the prior works' methodology~\cite{kim2020revisiting, orosa2021deeper, yaglikci2022understanding, luo2023rowpress}.}}

\begin{figure}[ht]
\centering
\includegraphics[width=\linewidth]{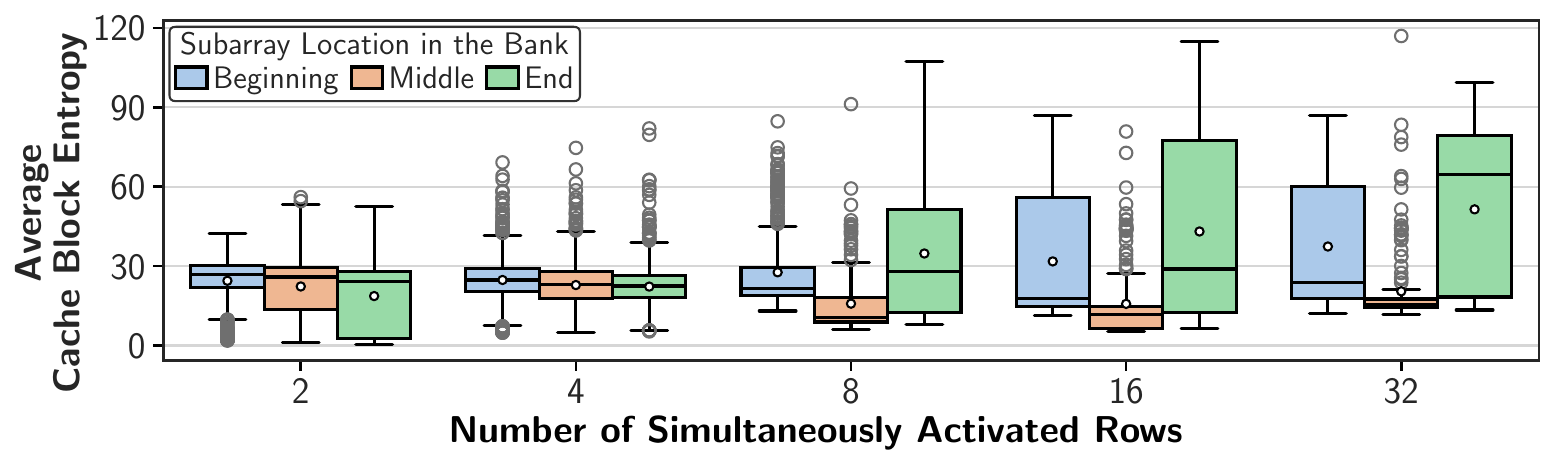}
\caption{Distribution of average cache block entropy based on the location of a DRAM subarray in a DRAM bank.}
\label{fig:spa_rows}
\end{figure}

\figref{fig:spa_rows} shows the distribution of average cache block entropy (y-axis) across {subarrays} for varying numbers of simultaneously activated rows (x-axis) based on the {subarray's} location in a {bank} (hue).
We make \textit{Observations 6 and 7} from \figref{fig:spa_rows}.

\obsv{The average cache block entropy significantly varies based on the {subarray's} location in a {bank}.}

We observe that the subarray's location in a bank can lead to variations in the average cache block entropy. \ieycr{0}{The average point of each bar's average cache block entropy (i.e., the average of each bar, white circles in \figref{fig:spa_rows}) varies} up to \param{1.31$\times$} for 2-row activation, \param{1.11$\times$} for 4-row activation, \param{2.20$\times$} for 8-row activation, \param{2.74$\times$} for 16-row activation, and \param{2.51$\times$} for 32-row activation.

\obsv{Each number of row activations has a different average cache block entropy variation trend.} 

We observe that in 2- and 4-row activations, when we get closer to the end of the bank, average cache block entropy consistently decreases. However, this trend is different for 8-, 16-, and 32-row activation: subarrays in the middle of the bank \omcr{0}{have} the lowest average entropy while subarrays at the end of the bank \omcr{0}{have} the highest average entropy. 

\takeawaybox{Average cache block entropy is significantly affected by spatial variation across DRAM subarrays.}

\section{SIMRA-TRNG}
\label{sec:simra_trng}
We demonstrate the potential of using simultaneous multiple-row activation (SiMRA) as the entropy source for a TRNG
by developing and evaluating a new TRNG design, SiMRA-TRNG.

\noindent\textbf{SiMRA-TRNG Design.} 
SiMRA-TRNG generates true random numbers \ieycr{0}{using four DRAM banks} in \param{five} steps: 
\one{} selecting a high-entropy \sar{} group based on our characterization data, 
\two{} initializing DRAM rows by performing the RowClone operation~\cite{seshadri2013rowclone, gao2019computedram, olgun2022pidram} in quick succession \ieycr{0}{in four DRAM banks}, 
\three{} performing \apa{} on the selected \sar{} to generate random bits in sense amplifiers \ieycr{0}{in four DRAM banks} \omcr{1}{concurrently}, 
\four{} reading from \ieycr{0}{four DRAM banks} until 256 bits of Shannon Entropy \omcr{0}{is obtained},\footnote{\label{fn:sha}We observe that 256-bit Shannon entropy is required as an input for the SHA-256 function for generating a 256-bit true random number.} 
and 5) post-processing the extracted bitstream using the SHA-256 cryptographic hash function~\cite{fips2012180} \ieycr{0}{to eliminate the bias and correlation and provide
protection against environmental changes and adversar\omcr{1}{ial} tampering~\cite{kocc2009cryptographic, rahman2014ti, stipvcevic2014true} (see \secref{subsec:trng})}.

\noindent\textbf{Quality.}
We evaluate the quality of SiMRA-TRNG using the NIST Statistical Test Suite (NIST STS) tests~\cite{rukhin2001statistical}. To maintain a reasonable testing time, we test random bitstreams from three DRAM modules, each from a different chip density and die revision pair. For each number of simultaneously activated rows (2-, 4-, 8-, 16-, and 32-row activation), we extract a 1 Gb bitstream from the highest-entropy \sar{} group, partition the bitstream into 1 Mb sequences, and test 1024 sequences per \sar{} group. 

\obsv{Random numbers generated with SiMRA and post-processed with the SHA-256 function pass \emph{all} NIST STS tests. We conclude that SiMRA-TRNG reliably produces high-quality true random bitstreams for all tested numbers of simultaneously activated rows.}

\noindent\textbf{Latency.}
We evaluate the latency of true random number generation by measuring the time required for three key operations:
\one{} performing \ieycr{0}{$N$} RowClone operations to initialize the \ieycr{0}{$N$} simultaneously activated rows \ieycr{0}{(e.g., for 2-row activation-based SiMRA-TRNG, we perform two RowClone operations)} \ieycr{0}{in four DRAM banks},
\two{} executing the \apa{} command sequence \iey{to activate a \sar{} group that has the highest entropy} \ieycr{0}{in four DRAM banks}, and
\three{} reading random values from the sense amplifiers and post-processing them with SHA-256.

\obsv{To generate first 256-bits of random numbers, the latencies for SiMRA-TRNG designs based on 2-, 4-, 8-, 16-, and 32-row activation are $\param{164.9}ns$, $\param{254.9}ns$, $\param{434.9}ns$, $\param{794.9}ns$, and $\param{1514.9}ns$, respectively.}

\noindent\textbf{Throughput.} 
We analytically model the throughput of SiMRA-TRNG for varying numbers of simultaneously activated rows based on two key factors:
\one{} $N_{block}$, the number of 256-bit Shannon entropy blocks available in the highest-entropy \sar{} group,\footref{fn:sha} and
\two{} $L_{TRNG}$, the latency of true random number generation \ieycr{0}{in four DRAM banks}. SiMRA-TRNG's throughput per DRAM module for each number of row activations is equal to ($256$$\times$$N_{block}$)$/$$L_{TRNG}$ bits per second.

\begin{figure}[ht]
    \centering
    \includegraphics[width=1\linewidth]{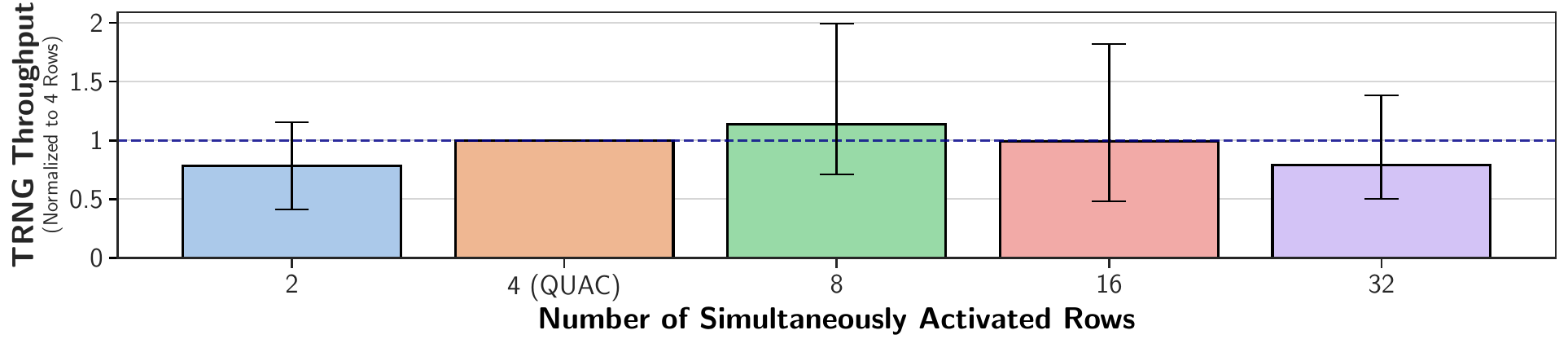}
    \caption{Throughput of generating true random numbers using SiMRA, normalized to state-of-the-art DRAM-based TRNG, QUAC-TRNG (i.e., 4-row activation)~\cite{olgun2021quactrng}.}
    \label{fig:mra_trng_results}
\end{figure}

\figref{fig:mra_trng_results} shows the throughput of SiMRA-TRNG, with error bars representing the range across all tested DRAM modules. Each module's throughput for varying numbers of activated rows is normalized to the state-of-the-art DRAM-based TRNG, QUAC-TRNG (i.e., 4-row activation)~\cite{olgun2021quactrng}. We make two key observations from \figref{fig:mra_trng_results}. 

\obsv{SiMRA-TRNG designs based on 2-, 8-, 16-, and 32-row activation outperform the state-of-the-art DRAM-based TRNG (QUAC-TRNG~\cite{olgun2021quactrng}\omcr{0}{, which uses} 4-row activation).} 

\omcr{0}{SiMRA with} 2-, 8-, 16-, and 32-row activations achieve up to $\param{1.15}$$\times$, $\param{1.99}$$\times$, $\param{1.82}$$\times$, and $\param{1.39}$$\times$ higher throughput than \omcr{0}{the} state-of-the-art, \omcr{0}{QUAC-TRNG~\cite{olgun2021quactrng},} respectively. We also observe that 2-, 4-, 8-, 16-, and 32-row activation-based SiMRA-TRNG designs achieve average throughputs of $\param{9.90}~Gbps$, $\param{13.81}~Gbps$, $\param{16.05}~Gbps$, $\param{13.94}~Gbps$, and $\param{10.81}~Gbps$, respectively. 

\obsv{There is a tradeoff between \ieycr{0}{TRNG} latency and the entropy produced by SiMRA.}

While simultaneously activating more rows tends to generate more entropy than activating fewer rows, it also increases the \ieycr{0}{TRNG} latency. For example, in all tested DRAM chips, 32-row activation achieves higher average cache block entropy than 16-row activation. 
However, the higher entropy observed with 32-row activation \textit{does not} lead to a higher throughput as the latency \ieycr{0}{of initializing the simultaneously activated rows} is also doubled compared to 16-row activation \ieycr{0}{(i.e., initializing 16 rows vs. 32 rows)}.

\takeawaybox{SiMRA-TRNG outperforms the state-of-the-art \omcr{0}{DRAM-based TRNG in terms of throughput} and offers a promising approach to generating true random numbers with high throughput.}

\section{Related Work}

This is the first work \ieycr{0}{to characterize the use of} the simultaneous multiple-row activation for varying numbers of simultaneously activated rows as a TRNG substrate in COTS DRAM chips.

\noindent\textbf{DRAM-based TRNGs~\cite{keller2014dynamic,sutar2018d,tehranipoor2016robust,eckert2017drng,kim2019d,talukder2019exploiting,pyo2009dram,olgun2021quactrng,bostanci2022drstrange}.} These works propose DRAM-based TRNGs using retention failures~\cite{keller2014dynamic, sutar2018d}, DRAM start-up values~\cite{tehranipoor2016robust, eckert2017drng}, timing failures~\cite{kim2019d,talukder2019exploiting}, the unpredictability in DRAM command schedule\omcr{0}{s}~\cite{pyo2009dram}.
The state-of-the-art DRAM-based TRNG~\cite{olgun2021quactrng} generates true random numbers in COTS DDR4 chips by simultaneously activating four rows. Our study \omcr{0}{generalizes this finding and} shows that simultaneously activating 2, 8, 16, and 32 rows generates \omcr{0}{high-quality} true random numbers and can achieve higher throughput than the state-of-the-art. \omcr{1}{DR-STRaNGe~\cite{bostanci2022drstrange} provides end-to-end system design for DRAM-based TRNGs.}

\noindent\textbf{SRAM- and FLASH-based TRNGs~\cite{vanderLeest2012,zhang2020improved,kiamehr2017leveraging,rahman2016enhancing,sadhu2020sstrng,wang2020aging,yeh2019self,holcomb2007initial,clark2018sram,yuksel2022turan,wang2020long,li2015pufkey,ray2018true,chakraborty2020true,wang2012flash}.} \ieycr{0}{SRAM-based TRNG designs exploit randomness in startup values~\cite{vanderLeest2012,zhang2020improved,kiamehr2017leveraging,rahman2016enhancing,sadhu2020sstrng,wang2020aging,yeh2019self,holcomb2007initial,clark2018sram,wang2020long,li2015pufkey} and randomness caused by reduced supply voltage when reading data~\cite{yuksel2022turan}. Flash-based TRNGs use random
telegraph noise in flash memory devices as an entropy source (up to 1 Mbit/s)~\cite{ray2018true,chakraborty2020true,wang2012flash}.}

\noindent\textbf{Multiple-Row Activation in COTS DRAM Chips~\cite{gao2019computedram,gao2022frac,olgun2021quactrng,olgun2022pidram,olgun2023dram,yuksel2023pulsar,yaglikci2022hira,yuksel2024functionally,yuksel2024simultaneous,yuksel2025pudhammer}.} Several \omcr{0}{experimental} works \omcr{0}{on real DRAM chips} demonstrate that COTS DRAM chips can activate more than one DRAM row simultaneously \omcr{0}{or sequentially in quick succession} \omcr{1}{by violating specific DRAM timing parameters}. These works show that COTS DRAM chips can \one{} perform \omcr{0}{bulk} bitwise operations (e.g., AND and NOT) operations~\cite{gao2019computedram,gao2022frac,olgun2023dram,yuksel2023pulsar,yuksel2024functionally, yuksel2024simultaneous}, \two{} copy one row's content \omcr{0}{in}to \omcr{0}{one or more} rows~\cite{gao2019computedram,olgun2022pidram,yuksel2023pulsar,yuksel2024simultaneous} and \three{} can concurrently refresh two rows in two subarrays~\cite{yaglikci2022hira}. \omcr{0}{These works~\cite{gao2019computedram,gao2022frac,olgun2021quactrng,olgun2022pidram,olgun2023dram,yuksel2023pulsar,yaglikci2022hira,yuksel2024functionally,yuksel2024simultaneous,yuksel2025pudhammer} are inspired by and based on RowClone~\cite{seshadri2013rowclone} and Ambit~\cite{seshadri2017ambit,seshadri2015fast}. RowClone~\cite{seshadri2013rowclone} and Ambit~\cite{seshadri2017ambit,seshadri2015fast} demonstrated using circuit theory, first principles, and simulation that DRAM chips can perform data copy \& data initialization and bulk bitwise operations via multiple-row activation. }

\noindent\textbf{\omcr{0}{Experimental Characterization and Analysis of COTS DRAM Chips.}}
\omcr{0}{Various prior works~\cite{orosa2021deeper,kim2020revisiting,kim2014flipping, lim2017active, park2016statistical, park2016experiments, ryu2017overcoming, yun2018study, lim2018study, luo2023rowpress, lang2023blaster, yaglikci2024svard, nam2024dramscope, olgun2023hbm,nam2023xray,luo2025revisiting,tugrul2025understanding,yuksel2025pudhammer,he2023whistleblower,luo2024experimental,olgun2025variable, girayphd,yuksel2025columndisturb,jung2015omitting,khan2014efficacy,khan2016parbor,khan2016case,kim2015architectural,liu2013experimental,meza2015revisiting,bianca-sigmetrics09,weis2015retention,patel2017reaper,gong2018memory,olgun2023dram,kim2018solar,yaglikci2022understanding,patel2019understanding,hassan2021utrr,patel2020beer,vampire2018ghose,chang2017understanding,lee2017design,chang2016understanding,kim2018dram,lee2015adaptive} analyze different phenomena in COTS DRAM chips using FPGA-based infrastructures like DRAM Bender~\cite{olgun2023dram,safari-drambender} \& SoftMC~\cite{hassan2017softmc,softmcgithub}. Understanding the interaction of these effects with SiMRA-TRNG or the use of these effects for better TRNG can be avenues for future work.}
\section{Conclusion}
We experimentally demonstrated and analyzed the potential \omcr{0}{and benefits} of using \omcr{0}{the} simultaneous multiple-row activation (SiMRA) \omcr{0}{phenomenon} in COTS DRAM chips to generate true random numbers. 
Through an extensive characterization using \nCHIPS{} DDR4 chips, we showed that \one{} using SiMRA as a TRNG substrate achieves higher throughput than the state-of-the-art \omcr{0}{DRAM-based TRNG} and \two{} the number of simultaneously activated rows, data patterns, temperatures, and spatial variations significantly affect the \gls{simra}'s entropy. \omcr{0}{To facilitate future research in using DRAM for novel purposes and discovering new capabilities of DRAM chips, we \omcr{1}{freely} open source our infrastructure at \url{https://github.com/CMU-SAFARI/SiMRA-TRNG}.}

\section*{Acknowledgments}
We thank the anonymous reviewers of DSN Disrupt 2024, VTS 2025, IMW 2025, and ICCD 2025 for their feedback. 
We thank the SAFARI Research Group members for providing a stimulating intellectual and scientific environment. We acknowledge the generous gifts from our industrial partners, including Google, Huawei, Intel, and Microsoft. This work{, and our broader work in Processing-in-Memory,} is supported in part by the Semiconductor Research Corporation (SRC), the ETH Future Computing Laboratory (EFCL), and the AI Chip Center for Emerging Smart Systems (ACCESS).

\balance
\bibliographystyle{unsrt}
\bibliography{refs}
\vfill

\end{document}